\documentstyle[12pt]{article}
\newcommand{\nc}[2]{\newcommand{#1}{#2}}
\newcommand{\Nc}[3]{\newcommand{#1}#2{#3}}
\Nc{\bib}[4]{\bibitem{#1} #2 \bf #3 \rm #4 }
\Nc{\me}[1]{\title{#1} \author{Mitsuru Ishii
\thanks{e-mail:\ ishii@jpnyitp.bitnet} } \date{} \maketitle \par
\noindent 
\centerline{Yukawa Institute for Theoretical Physics, Kyoto, 606-01, Japan}}
\Nc{\up}[2]{\stackrel{#1}{#2}}
\Nc{\bo}[1]{\vskip 3mm\centerline{\fbox{#1}}\vskip3mm}
\Nc{\appe}[1]{\section*{Appendix #1}

\setcounter{equation}{0}}
\nc{\eqnoforeachsection}{}
\nc{\enb}{\begin{enumerate}}
\nc{\ene}{\end{enumerate}} 
\nc{\eq}{\begin{eqnarray}}
\nc{\eqe}{\end{eqnarray}} 
    \nc{\be}{      
\begin{document}}
\nc{\bye}{\end{document}}
\nc{\ab}{\begin{abstract}}
\nc{\abe}{\end{abstract}}
\nc{\itb}{\begin{itemize}}
\nc{\ite}{\end{itemize}}
\nc{\bibb}{}
\nc{\non}{\nonumber}
\nc{\nopage}{\pagestyle{empty}}
\nc{\nextpage}{\newpage}
%\be
%
%
\nc{\bra}{\langle}
\nc{\ket}{\rangle}
\nc{\rla}{\longrightarrow}
\nc{\ra}{\rightarrow}
\nc{\gcon}{\la G^2 \ra}
\nc{\reac}{$\pi+\pi\leftrightarrow\pi+\pi+\pi+\pi$} 
\nc{\scat}{$\pi+\pi\leftrightarrow\pi+\pi$} 
\be
\me{Nonequilibrium process in the $\sigma$ model and chemical
relaxation time in a homogeneous pionic gas}
%\maketitle
%\vskip-10cm\rightline{YITP-96-32}\vskip10cm

\abstract{In a homogeneous pionic gas system, chemical nonequilibrium
process is considered to understand its effect in the expansion 
processes that are realized immediately after heavy ion collisions.
The chemical relaxation time is calculated by incorporating the 
$\pi+\pi \leftrightarrow \pi+\pi+\pi+\pi$ reaction, which is
given in the second order of perturbation in the $\sigma$ model.
The $\pi+\pi \leftrightarrow \pi+\pi+\pi+\pi$ reaction is assumed to
be less frequent than the $\pi+\pi \leftrightarrow \pi+\pi$ scattering 
that is expected to establish the local equilibrium, and
hydrodynamical equation is solved for various initial conditions.
It is shown that the relaxation time is of the order of 100fm
and does not have a significant effect on the expansion process,
which implies that the pion number freezeout takes place at an early
stage of the expansion.}

\newpage
\section{Introduction}
\indent

 It is widely believed that Quantum Chromo Dynamics (QCD)
is the underlying theory of strong interaction and that QCD exhibits 
confinement-deconfinement phase transition at high temperature
($\approx $200 MeV{\cite{lattice}.) Since understanding the property of
this phase transition is crucial for hadron physics, many efforts have
been made both experimentally and theoretically \cite{the}.
In particular, heavy ion reaction is
expected to be a useful method to establish high temperature with free
quarks and gluons and to investigate the process of hadronization.

Heavy ion collision is considered to undergo several steps shown in Fig.1.:
\enb
\item Two highly Lorentz contracted nuclei collide with each other. 
\item They pass through each other and a hot system is formed in between,
where quarks and gluons are excited\cite{rev}.
\item Frequent collisions and reactions among the excited elements
lead to establishing local equilibrium. Therefore this state can be 
treated by hydrodynamics\cite{mat} \cite{berndt}.
\item As the system expands, cooling and subsequent phase transition
process proceeds\cite{the}.
\item A lot of pions ( and other light hadrons) are created under 
the process of the phase transition\cite{bertsch}.
\item When the phase transition terminates, there exists hadronic 
gas which can also be treated by hydrodynamics.
Since the system is hot, the collisions and reactions between the
hadrons are frequent and the total number of hadrons is not conserved.
\item As the cooling goes on, the reaction to change the
total number of hadrons becomes unlikely. At this stage, the multiplicity
of pion (and other light hadrons) is fixed (freezeout of number)\cite{freeze}.
\label{freezeout}
\item When the expansion proceeds further and the mean distance
between pions is equivalent to the average variation of macroscopic
quantity such as temperature, energy density and pressure, the
hydrodynamical picture is not a good one any longer. Instead, the
system can be considered to be composed of free pions without any
interaction. This stage is called the freezeout of temperature. 
\label{tempfreeze}
\ene
{\bo{Fig.1}}

Therefore, the multiplicity of pion is determined at stage 
\ref{freezeout} not at the time when the phase transition terminates.
Also, the energy distribution of pions is given by the thermal 
distribution at stage \ref{tempfreeze}.
In other words, hadronic final state of heavy ion collision does not 
directly reflect the property of the confinement-deconfinement phase
transition 

Due to this feature, it has been proposed to use 
electromagnetic and weak probe\cite{asa},
which is free from the hadronic final state interaction, to
investigate the hadronic property of the phase transition. However, if
we can understand the process of the final state interaction, it is
possible to extract the information on the phase transition
by hadronic probes.
In particular, pion multiplicity can be used to obtain the pion number
and the entropy immediately after the phase transition: 
if one knows the elementary chemical process to change the 
pion multiplicity such
as \reac, one can see how the the pion multiplicity varies through the 
cooling process, and from the information on the pion multiplicity
obtained in  heavy ion reactions one can track the process backward up
to the time when the phase transition is over.

Whether the chemical reaction largely affects the cooling
process or not depends on the balance between the cooling
by the expansion and the speed of the chemical reaction.
If the typical time scale of the chemical reaction is longer enough than
that of the expansion, the freezeout of pion multiplicity
takes place at an early stage of the expansion and one can identify
the final pion multiplicity with the pion multiplicity immediately
after the phase transition in the first approximation.
In this case we can estimate the entropy density immediately after
the phase transition ($s$) by 
\eq
\label{entropy}
s=4 n,\ \ \ \   n:{\hbox{observed pion multiplicity}},
\eqe
which is obtained for massless classical particle systems with the
particle numbers conserved\cite{landau}.
If the former is shorter than the latter, the chemical reaction remains
dominant through the expansion. In this case, the final pion multiplicity
is quite different from the initial one at the end of the phase
transition, and Eq.(\ref{entropy}) can not be used to obtain the
entropy density. Instead, one must track the irreversible process
backward, estimate the pion multiplicity at the end of the phase
transition and then use Eq.(\ref{entropy}).

 In this paper, we aim to obtain the speed of the chemical reaction 
\reac\  based on the $\sigma$ model. 
In this model, the interaction in
the leading order of perturbation just gives 
$\pi+\pi\leftrightarrow \pi+\pi$ scattering, which is expected to
establish local equilibrium in a pionic gas. In the second
order of the perturbation, \reac\  reaction is derived. If the latter
reaction is less frequent than the former one, we can resort to
hydrodynamical equation to track the change of pion number
multiplicity. In section \ref{evalfreq}, the evaluation 
of the frequency of these 
two reaction is made and the criterion to establish local equilibrium
is given.
Then, we present hydrodynamical equation
in the classical approximation.
In section \ref{homog}, the hydrodynamical equation is solved
numerically in a homogeneous pion gas 
for various initial temperatures and chemical potentials.
The typical time scale of chemical reaction (relaxation time) 
is obtained and compared to the expansion rate.
%In Section \ref{expan}, we taken expansion into account 
%under several reasonable assumptions. It is confirmed that
%the exothermic chemical reaction plays an important role in the
%time evolution of temperature and chemical potential.
The last section \ref{discu} is devoted to summary
and discussing open problems.

\section{Elementary chemical reaction in the $\sigma$-model and 
the hydrodynamical equation}
\label{evalfreq}
\indent

We take $\sigma$-model to describe the \reac\  reaction.
The interaction part of the Lagrangian is given by
\eq
{\cal L}_{\rm int}=%{1 \over 2}\partial_\mu \sigma \partial^\mu \sigma 
        %+{1 \over 2}\partial_\mu {\bf \pi} \partial^\mu {\bf \pi}
        -{\lambda \over 4!}(\sigma^2+{\bf \pi}^2-{f_\pi}^2)^2,
\eqe
where $\pi=\pi^1\tau_1+\pi^2\tau_2+\pi^3\tau_3$ and $f_\pi$ is the
pion decay constant.
After shifting the $\sigma$ field as $\sigma\rightarrow\sigma+f_\pi$,
  one obtains $-\lambda/4! ({\bf \pi}^2+2 f_\pi\sigma+\sigma^2)^2$
as the potential term. This term gives \scat\  scattering depicted in
Fig.2(a). Summing all the diagrams in Fig.2(a) at threshold
$p_i=(m,0,0,0),\ (i=1,2,3,4)$ gives
\eq
-\lambda +({\lambda\over 3})^2{{f_\pi}^2\over
{m_\sigma}^2}+({\lambda\over 3})^2{{f_\pi}^2\over
{m_\sigma}^2}+({\lambda\over 3})^2{{f_\pi}^2\over {m_\sigma}^2-(2 m)^2},
\eqe
where $m_{\sigma}\ (m)$ is the $\sigma$ ($\pi$) meson mass.
Since ${m_\sigma}^2=\lambda {f_\pi}^2/3$, we get
\eq
-{\lambda\over 3}+{\lambda\over 3} {1\over {1-({2 m\over m_\sigma})^2}}
\eqe
as an  effective $4\pi$ coupling constant. For $m_\sigma\approx$ 600
MeV, $m/m_\sigma\approx1/4$ and the effective coupling constant is
$\lambda/9$. For simplicity, we hereafter replace the sum of Fig.1(a)
by just one direct $4\pi$ coupling with its coupling constant
$\lambda/9$.

The \reac\  reaction is given in the second order of perturbation
(Fig.2(b)). In the second order, another diagram is possible for 
the \reac\  process (Fig.2(c)). This diagram, however, is not taken 
into account because the denominator of the propagator
$p^2-m^2$ is of the order of $(3m)^2-m^2$ and the
contribution of Fig.2(c) is $O(1/10)$ of Fig.2(b).
All the possible diagrams of type (b) are shown in Fig.2(d). For simplicity,
we replace Fig2.(d) by Fig2.(e), which means the effect of
interference is not exactly taken into account.

\bo{Fig.2}

 In the next section, the hydrodynamical equation is
solved assuming local equilibrium that is expected to be established by
the \scat\  scattering. Since \reac\  reaction drives the system away from
equilibrium, \scat\  scattering must be frequent enough so that 
the disturbed nonequilibrium state can swiftly return to equilibrium
state.

Once local equilibrium is established, local temperature $T(x)$ as well as
local chemical potential $\mu(x)$ is defined. The \scat\  actually
means $\pi^0+\pi^0\leftrightarrow \pi^++\pi^-$ and
$\pi^i+\pi^j\leftrightarrow \pi^i+\pi^j\  (i,j=+,-,0)$.
The first one gives the equilibrium condition
$2\mu^0(x)=\mu^+(x)+\mu^-(x)$. Since the charge conservation
means $\mu^+(x)=\mu^-(x)$ in neutral pionic gas, we get
$\mu^+(x)=\mu^-(x)=\mu^0(x)$ or $n^{\pi^+}(x)=n^{\pi^-}(x)=n^{\pi^0}(x)$
($n$: pion density.) 
This means that all the equations are identical for $\pi^0,\pi^+$ and $\pi^-$.

In other words, all the calculation is made in the same way as in
$\lambda \phi^4$ theory except for a numerical factor stemming 
from the isospin degrees of freedom,
because the essential part of the $\sigma$
model is only the $(\lambda/4!) {\bf \pi}^4$ term. 

 As we have mentioned before, the \scat\  scattering should be more frequent
than \reac\  reaction in order to have the system locally equilibrated.
The frequency of \scat\  and \reac\  process are calculated in Appendix A:
\begin{eqnarray}
%setcounter
I^{\pi+\pi\leftrightarrow\pi+\pi}
&\equiv & \int {d{\bf p}_1 \over (2\pi)^3 2 {p_1}^0}
{d{\bf p}_2 \over (2\pi)^3 2 {p_2}^0} 
{d{\bf p}_3 \over (2\pi)^3 2 {p_3}^0}
{d{\bf p}_4 \over (2\pi)^3 2 {p_4}^0}\non\\
& &\times({\lambda\over 9})^2 (2\pi)^4 \delta(p_1+p_2-p_3-p_4)
e^{-\beta(p_1+p_2)^0+2\beta \mu}\non\\
&= &
({\lambda\over 9})^2 {1\over {32\ (2\pi)^5}} e^{2 \beta \mu} 
\int_{2m}^\infty dP_0 e^{-\beta P_0}\non\\
& &\times \int_0^{\sqrt{P_0^2-4 m^2}} d|{\bf p}| |{\bf p}|^2 
\sqrt{1-{4 m^2 \over (P_0^2-{\bf p}^2)}}
\end{eqnarray}
\eq
\label{ai}
I={575\over 243}
\ 3 \ ({\lambda\over 9})^4{1\over {128 (2\pi)^9}} e^{-4\beta} m^4
(e^{2\beta \mu}-e^{4 \beta \mu})
\int_0^\infty x^2 (x+8) g(x) e^{-\beta x} dx%\eqno{\ref{reac}},
\non\\
\eqe
where $\beta=1/T$, $T$ is the temperature, $\mu$ the chemical
potential, $m$ the pion mass,
$P_0$ the total energy of the colliding pions, ${\bf p}$
their total spatial momentum, $P_0/m=4+x$ and $g(x)$ a function
independent of temperature and chemical potential whose definition is
given in Appendix A. The factor $575\over 243$ comes from the 
isospin degrees of freedom (see Appendix B.)
Since no further analytical calculation is possible,
we carried out the integrals numerically for $m\beta\approx 1$
and $\mu/m\approx 1$ to get a rough estimate for 
$I^{\pi+\pi \leftrightarrow \pi+\pi}$ and $I$.
It was found that $\lambda \ll 1000$ is required for 
$I^{\pi+\pi\leftrightarrow\pi+\pi}\gg I$. Since $\lambda\approx 100$
is taken for the ordinary $\sigma$-model, local equilibrium
is established in the present case.\footnote{
Since \scat\  is $O(\lambda^2)$ and \reac\  is $O(\lambda^4)$, it may seem
that $\lambda\ll 1$ is required for local equilibrium. However, 
numerical factors coming from the phase space
integral such as $1/(2\pi)^3$ for each external line
and $e^{-4 \beta}$ in Eq.(\ref{ai})
suppress the frequency of the \reac\ process. Thus $\lambda\approx 100$
satisfies the criterion for local equilibrium even if $\lambda$ is
bigger than 1.}

 Under local equilibrium, relativistic hydrodynamical 
equation{\cite{deg}} is given as :
\begin{eqnarray}
\label{hydroori1}
%\left{
\partial_\mu {N^i}^\mu&=&2 I\\
\label{hydroori2}
\partial_\mu {T^i}^{\mu \nu}&=&0
%\right
\end{eqnarray}
for each species of pion, where ${N^i}^\mu (i=+,-,0)$ is pion number 
current and ${T^i}^{\mu \nu}$
is the energy momentum tensor for $\pi^+,\pi^-$ and $\pi^0$ respectively.
As is mentioned above, these fundamental equations take the same form
for all the species of pions. Hence, we hereafter suppress the
indices $i=+,-,0$. Eq.({\ref{hydroori1}}) describes
the change of the total pion number and is called rate equation,
whose right hand side $2I$ gives the rate of particle production
and annihilation. The factor 2 in the right hand side comes from the
fact that two pions are created for one elementary reaction
in Fig.2(b). Eq.(\ref{hydroori2}) is the energy-momentum conservation
identity.

$N^\mu$ and $T^{\mu \nu}$ are expressed in terms of the pion
distribution function $f$:
\eq
N^\mu=\int {d{\bf p} \over (2\pi)^3} {p^\mu \over p^0} f
\eqe
\eq
T^{\mu \nu}=\int {d{\bf p} \over (2\pi)^3} {{p^\mu p^\nu}\over p^0} f.
\eqe

\section{Numerical solution for the hydrodynamical equation and chemical
relaxation time in a homogeneous pionic gas} 
\label{homog}
\indent

In this section, we consider homogeneous pionic gases to obtain the
typical time scale of the chemical reaction \scat. In a homogeneous
gas, the hydrodynamical equation Eqs.(\ref{hydroori1})(\ref{hydroori2}) is
drastically simplified to\cite{mat}
\eq
\label{homorate}
{{d n} \over {d t}}= 2 I
\eqe
\eq
\label{homoene}
{{d \epsilon}\over {d t}}=0,
\eqe
where $n=N^0 (\epsilon=T^{00})$ is the pion density (energy density).
To express $N^\mu$ and $T^{\mu \nu}$ in terms of the local temperature
$T$ and  local chemical potential $\mu$,
we need to have the explicit form of the distribution
function $f$.
In this paper, we take classical distribution
\eq
f(p,x,t)={\rm exp}\{-\beta(x,t) (\sqrt{{\bf p}^2+m^2}-\mu(x,t))\}.
\eqe
Since we investigate homogeneous pionic gas, there is no $x$
dependence in any macroscopic quantity and $T$ and $\mu$ are the functions
of time $t$ alone.
Then, we get
\eq
\label{numden}
n(t)= {1\over 2\pi^2} e^{\mu/T} T m^2 K_2({m\over T})
\eqe
\eq
\label{eneden}
\epsilon(t) ={1\over 2\pi^2} e^{\mu/T} m^2 \{3 T^2 K_2({m\over T})
+T m K_1({m \over T})\}
\eqe
Combining Eqs.(\ref{homorate})(\ref{homoene}) with
Eqs.(\ref{numden})(\ref{eneden}), one gets a set of differential
equation for $T(t)$ and $\mu(t)$. The energy conservation law 
Eq.(\ref{homoene}) means $\epsilon(t)=\epsilon(t=0)$ giving the
equation to determine $\mu(t)$ from $T(t)$ and initial conditions
$T_0=T(0),\mu_0=\mu(0)$:
\eq
\label{repmu}
\mu={\mu_0\over T_0} T+T \ln \{({T_0\over T})^2 
    {{3 K_2(m/T_0)+m/T_0 K_1(m/T_0)} \over 
    {3 K_2(m/T)+m/T K_1(m/T)} }\},
\eqe
with the abbreviation $\mu=\mu(t),T=T(t)$.
Also from $d\epsilon/dt=0$, we can express $d\mu/dt$ in terms of
$T(t),\mu(t),dT/dt,T_0$ and $\mu_0$:
\eq
{\dot \mu}=({\mu\over T}-A){\dot T}
\eqe
\eq
A\equiv {{12 K_2(m/T)+5 (m/T) K_1(m/T)+(m^2/T^2) K_0(m/T)}\over
          {3 K_2(m/T)+(m/T)K_1(m/T)}},
\eqe
where ${\dot \mu}\  ({\dot T})$ means $d\mu/dt\  (dT/dt)$.
By eliminating $\mu$ and $\dot \mu$ and redefining all the quantities
to be dimensionless, i.e. 
$T/m\rightarrow T\equiv1/\beta,\mu/m\rightarrow
\mu,\lambda^4mt\rightarrow t$, 
 we can rewrite the hydrodynamical equation as
\eq
\{(3-A)K_2(\beta) +\beta K_1(\beta)\}{\dot T}= 2 I/\lambda^4.
\eqe
Since $I$ is proportional to $\lambda^4$ (see Eq.(\ref{ai})),
the $\lambda^4$ term in 
the redefinition of $t$ makes the equation independent of $\lambda$.

By solving this differential equation, one gets the $t$ dependence of
$T$ and from Eq.(\ref{repmu}) $\mu(t)$ is also obtained.
The results of numerical calculation are shown in Figs.3,4,5,6
for initial conditions $(T_0,\mu_0)=(0.5,-1),(1.5,-1),(0.5,1),(1.5,1)$
respectively. 
\bo{Figs.3,4,5,6}
From these figures, one can see that the relaxation is
faster for hotter and denser (larger $\mu_0$) systems, which is
reasonable considering that the chemical reaction is more frequent
for hotter and denser systems. Since the relaxation is not exactly
exponential, we define relaxation time by $\mu(t_{\rm relax})=\mu_0/e$.
The relaxation times with $\lambda=120$ 
are shown in Fig.7 for various initial conditions
$-1<\mu_0<1, 0.5<T_0<1.5$.
For a dense and hot system (e.g. $\mu_0=1$ and $T_0=1.5$), 
$t_{\rm relax}$ is of the order of $100$fm.
This time scale is
expected to be larger than the typical time scale of expansion.
Thus, we have found that the \reac\  reaction does not play an important role
in an expanding pionic gas system.
In other words, the pion number freezes out at an early stage of the
final state of a heavy ion reaction. This implies that the hadronic
probe is a good one to extract information on the phase transition
and that $s=4 n$ is an appropriate one to evaluate the entropy density
immediately after the phase transition,  

\bo{Fig.7}
\section{Summary and open problems}
\label{discu}
\indent

 We have investigated the effect of \reac\ chemical reaction in the
framework of hydrodynamics. To understand the typical time scale
of the \reac\ reaction, we calculated the relaxation times in
homogeneous pionic gas systems. It turned out that the relaxation
time is of the order of 100fm at shortest and is longer than
typical time scale of the expansion of the pionic gas. 

Although we found that the chemical \reac\  reaction does not play an
important role within the framework of the $\sigma$-model, 
there still are a lot of things to be done.
%possibilities that the total pion number
%changes through the process of expansion.
The first one is to take into account the effect of the baryon in the central
region, which has been neglected so far.
In the real case, thermally excited $N\bar{N}$ pairs exist in the
central region. Since the $\pi N$ interaction is much larger than
the $\pi \pi$ interaction, the processes such as $N+\pi\leftrightarrow
N+\pi+\pi$ might be significant even if the number of the excited 
$N\bar{N}$ pairs is small. Furthermore, if there is a stopping in
heavy ion collisions and there exists baryon number coming from
the colliding nuclei, there would be a big change of pion multiplicity.   

% extend our method to expanding pionic gas systems and quantitatively
%investigate the balance between expansion and \reac\ reaction.
Secondly, to make reliable quantitative discussion on the role of the
\reac\ reaction, the calculation must be made without classical
approximation incorporating the effect of interference.
This is expected to make the collision less frequent and lead to
a larger relaxation time, keeping the essential statement in the
former section.
The third thing is to take the medium effect into account. 
So far, we have fixed the parameters in the $\sigma$-model:
the coupling constant $\lambda$ and the pion mass $m$.
It is known, however, that these quantities depend on temperature
and chemical potential. The correction due to this effect might not be
negligible.
%Last but not least, the ideal pionic gas without nucleon may not
%be the real case in  heavy ion collision. There is a possibility
%that stopping takes place in heavy ion collision, which means
%the existence of baryon number in the central rapidity region.
%Since the interaction between quark and gluon is much stronger than
%the $\pi$-$\pi$ interaction, this could lead to a big modification
%to the results in this paper. Even if there is no baryon number,
%$N{\bar N}$ can be produced at high temperature, which again
%implies unnegligible effect.

Even  if some of these effects lead to a smaller relaxation time and
make the effect of the change of the pion number a little more
important, we still do not expect them to change our results
qualitatively because of the following reason. 
In an expanding system, the chemical potential at the end 
of the phase transition is taken to be 0, because there are strong
interaction between QGP and pions and pion can change its multiplicity
easily. Therefore, at an early stage of the
expansion in which the system is hot enough, the chemical potential
is close to 0 and the system is close to chemical equilibrium.
So, \reac\ reaction etc. can not play an important role. As the system
expands, the chemical potential deviates from 0. However, the system
is not hot enough and the relaxation time is too long to make the
effect of \reac\ reaction significant.

%We do not think that these effects change the qualitative
%feature in this paper
% but they are important in order to give 
%quantitative results.

\section*{Acknowledgement}
\indent

The author would like to thank Prof.T.Matsui for fruitful discussions
with him and also thanks to members of nuclear theory group at
Kyoto University for their giving useful comments.
\newpage
\appe{A}

In this appendix, we show how to calculate the frequency of 
\scat \ scattering and \reac\ reaction. For simplicity, we
neglect the isospin degrees of freedom, which will be taken
into account in Appendix B.

Under classical approximation, the frequency of the process depicted
in Fig.2(a) per unit volume and unit time is given by
\eq
\label{iscatt}
I^{\pi+\pi\leftrightarrow\pi+\pi}
&= & \int {d{\bf p}_1 \over (2\pi)^3 2 {p_1}^0}
{d{\bf p}_2 \over (2\pi)^3 2 {p_2}^0} 
{d{\bf p}_3 \over (2\pi)^3 2 {p_3}^0}
{d{\bf p}_4 \over (2\pi)^3 2 {p_4}^0}\non\\
&\times&({\lambda\over 9})^2 (2\pi)^4 \delta(p_1+p_2-p_3-p_4)
e^{-\beta(p_1+p_2)^0+2\beta \mu}.
%\eqno{(A.1)}
\eqe
We first carry out the integral over $d{\it\bf p}_3$ and 
$d{\it\bf p}_4$.
Rewriting the integral in an explicitly Lorentz invariant form by
using 
\eq
\label{lorentz}
\int {d{\bf p}_3 \over (2\pi)^3 2 {p_3}^0}=\int dp_3 \ \theta(p_3^0)
\ \delta(p_3^2-m^2)
%\eqno{(A.2)}
\eqe
and then changing the variables to
$p=p_3+p_4,q={p_3-p_4\over 2}$,
we get
\eq
& &\int {d{\bf p}_3 \over (2\pi)^3 2 {p_3}^0} 
{d{\bf p}_4 \over (2\pi)^3 2 {p_4}^0}
\ \delta(p_3+p_4-p_1-p_2)\non\\
&=& \int dp_0d{\bf p} \int_{-{|{\bf p}|\over 2}\sqrt{1-{4 m^2\over
p^2}}}^{{|{\bf p}|\over 2}\sqrt{1-{4 m^2\over p^2}}}
dq_0 {1\over {|{\bf q}||{\bf p}|}} 2\pi^2 \int d|{\bf q}||{\bf q}|^2
{1\over 2}\delta(p-p_1-p_2) \non\\
&\times&
\int d(\cos\theta)
\delta(\cos\theta-{{p^0 q^0}\over {|{\bf p}| |{\bf q}|}})
\delta(|{\bf q}|-\sqrt{q_0^2+{p^2\over 4}-m^2})
{1\over {2 \sqrt{q_0^2+{p^2\over 4}-m^2}}}\non\\
&=&{\pi \over 2}\sqrt{1-{{4 m^2}\over (p_1+p_2)^2}}.
%\eqno{(A.4)}
\eqe
%where $\theta_1$ is the angle between ${\bf P}$ and ${\bf Q}$,
%$\theta_2$ ${\bf P}$ and ${\bf q}_t$.
The integral over $d{\bf p}_1$ and $d{\bf p}_2$ is made 
essentially in the same way.
However, the factor $\sqrt{1-{{4 m^2}\over (p_1+p_2)^2}}$
in the above equation now prevents full analytical calculation.
The final result is
\eq
I^{\pi+\pi\leftrightarrow\pi+\pi}&=&
({\lambda\over 9})^2 {1\over 32\ (2\pi)^5} e^{2 \beta \mu}
\int_{2m}^\infty d{P^\prime}_0 e^{-\beta {P^\prime}_0}\non\\
& &\times \int_0^{\sqrt{{P_0^\prime}^2-4 m^2}} d|{\bf p}| |{\bf p}|^2 
\sqrt{1-{4 m^2 \over ({P_0^\prime}^2-{\bf p}^2)}},
\eqe
where ${P_0^\prime}=(p_1+p_2)^0$ and ${\bf p}={\bf p}_1+{\bf p}_2$.

The collision frequency for Fig.2(b) is made almost in the same
manner. All the possible diagram of type Fig2.(b) is given 
in Fig2.(d). We approximate Fig2(d) by Fig2.(e), which means that
we have replaced the interference term by Fig.2(e).
This approximation makes it possible to carry out the integral over
the relative
variables between $p_1$ and $p_2$ and between $p_3$ and $p_4$.
%Now, the propagator prevents analytical 
%calculations and we get
The rate of the particle production of the processes
$\pi+\pi\leftarrow \pi+\pi+\pi+\pi$ depicted in Fig.2(e)is written as
\eq
& &{1\over 4!2!} \int
{d{\bf p}_1 \over (2\pi)^3 2 {p_1}^0} 
{d{\bf p}_2 \over (2\pi)^3 2 {p_2}^0} 
{d{\bf p}_3 \over (2\pi)^3 2 {p_3}^0}
{d{\bf p}_4 \over (2\pi)^3 2 {p_4}^0} 
{d{\bf p}_5 \over (2\pi)^3 2 {p_5}^0}
{d{\bf p}_6 \over (2\pi)^3 2 {p_6}^0}\non\\
&\times& e^{2\beta \mu} e^{-\beta (p_5+p_6)^0}
|6 M|^2 (2\pi)^4\delta(p_1+p_2+p_3+p_4-p_5-p_6),
\eqe
where $M$ is the matrix element of the Fig.2(b):
\eq
M=({\lambda\over 9})^2 {1\over m^2-(p_1+p_2-p_5)^2}.
\eqe
The rate for the reverse reaction $\pi+\pi+\pi+\pi\leftarrow \pi+\pi$
is given by replacing
$e^{2\beta \mu}e^{-\beta (p_5+p_6)^0}$ by 
$e^{4\beta \mu}e^{-\beta (p_1+p_2+p_3+p_4)^0}$.
%and flipping the sign.
The difference from Eq.(\ref{iscatt}) is that the propagator in $M$
hampers analytically carrying out the integral of $d(p_1+p_2)$ and
$d(p_3+p_4)$ as is shown below.

Taking $p_1+p_2=p,{p_1-p_2\over 2}=q,
p_3+p_4=p^\prime,{p_3-p_4\over 2}=q^\prime$ and carrying out
the integral over $dq$ and $dq^\prime$ as is done in
Eq.(\ref{lorentz}), we get
\eq 
& &{3\over 4} e^{2\beta \mu} {1\over (2\pi)^{18}}({\pi \over 2})^2\non\\
& &\times \int_{p^2>4 m^2}dp \sqrt{1-{4 m^2\over p^2}}
\int_{{p^\prime}^2>4 m^2}d{p^\prime} \sqrt{1-{4 m^2\over
{p^\prime}^2}}\int d p_5 d p_6 \non\\
& &\times \delta({p_5}^2-m^2) \delta({p_6}^2-m^2) \theta({p_5}^0) \theta({p_6}^0)
\delta(p+p^\prime+p_5+p_6)\non\\
& & \times e^{-\beta (p_5+p_6)^0} \{ {1\over {m^2-(p-p_5)^2}} \}^2.
\eqe
Then, we rewrite $p_5+p_6=p_t, {{p_5-p_6}\over 2}=q_t,
p+p^\prime=P,{p-p^\prime\over 2}=Q$ and define
\eq
P&=&(P^0,0,0,|{\bf P}|)\non\\
Q&=&(Q_0,|{\bf Q}|\sin\theta_1,0,|{\bf Q}|\cos\theta_1)\non\\
q_t&=&({q_t}^0,|{\bf q}|\sin\theta_2\cos\varphi,
|{\bf q}|\sin\theta_2\sin\varphi,|{\bf q}|\cos\theta_2) .
\eqe
$p_t$ can be eliminated due to the total momentum conservation 
$\delta$ function. The other two $\delta$ functions are written as
\eq
\delta({p_5}^2-m^2) \ \delta({p_5}^2-m^2)
= \delta({P^2\over 4}+Q^2-m^2) \ \delta(2 P\cdot Q)
\eqe
and enable us to eliminate $\cos\theta_2$ and $|{\bf q}_t|$.
The integration over $d\varphi$ is easily made and after some
manipulation to make the expression convenient for Monte Carlo
calculation, we get 
%Then the number of produced pion per unit time and unit volume is
%given by
\eq
I&=&{3\over {128 (2\pi)^9}} ({\lambda\over 9})^4e^{-4\beta} m^4
(e^{2\beta \mu}-e^{4 \beta \mu})\non\\
& &\times \int_0^\infty x^2 (x+8) g(x) e^{-\beta x} dx,
\eqe
\eq
g(x)=& &\int_0^1dx_2 \int_{-1}^1dx_3 \int_0^1dx_4
\int_{-1}^1dz \int_{-1}^1dy\non\\
&\times&
l^2 \ \sqrt{(1-{4 \over {{P^2\over 4}+Q^2+P\cdot Q}})
(1-{{P^2\over 4}+Q^2-P\cdot Q})}
{|{\bf P}|\over 2}\sqrt{1-{4 \over P^2}}{a\over \sqrt{a^2-b^2}}\non\\
&\times&
\theta({P^2\over 4}+Q^2-4-|P\cdot  Q|)\  x_2\  x_4^2
\eqe
with dimensionless variables redefined by
\eq
& &P={{p_1+p_2+p_3+p_4}\over m}={{p_5+p_6}\over m},
Q ={{(p_1+p_2-p_3-p_4)}\over 2 m},
q_t={{(p_5-p_6)\over 2 m}},\non\\
& &{P^0}=4+x, {|{\bf P}|}=\sqrt{{{P^0}^2}-16}x_2
=\sqrt{x (x+8)} x_2,
{Q_0}={x\over 2} x_3,|{\bf Q}|=l x_4,\non\\
& &{q_t}^0={|{\bf P}|\over 2}\sqrt{1-{4\over P^2}} y,
 |{\bf q}_t|=\sqrt{{{P_0}^2-|{\bf P}|^2\over 4}-1+({q_t}^0)^2},\non\\
& &l=\sqrt{{{P^0}^2-16\over 4}+{Q^0}^2},
a={P^2\over 4}-Q^2+2Q^0{q_t}^0-2|{\bf Q}||{\bf q}_t|z_1 z_2,\non\\
& & b=2|{\bf Q}||{\bf q}_t|\sqrt{1-z_1^2}\sqrt{1-z_2^2},
z=z_1=\cos\theta_1,z_2=\cos\theta_2={P^0 {q_t}^0\over |{\bf p}||{\bf q}_t}|.
\eqe
%$\theta_1$ is the angle between {\bf P} and {\bf Q}, and 
%$\theta_2$ between {\bf P} and ${\bf q}_t$.
The $\theta$ function comes from the constraints
\eq
{p}^2=({P\over 2}+Q)^2>4m^2\ \ \hbox{and}\ \ {p^\prime}^2=({P\over 2}-Q)^2>4m^2
\eqe
The advantage of taking the classical distribution is that the thermal
factor is given as $e^{-\beta P_0}$ and all the variable other than
the total energy can be integrated regardless of temperature.
The temperature independent part is written as $g(x)$ whose $x$  (or
$P_0$) dependence is shown in Fig.A.
\bo{Fig.A}
%\newpage
\appe{B}

\par\indent 

In this appendix, we consider the effect of the isospin degrees of
freedom. For simplicity, we take $\pi^0$,$\pi^1$,$\pi^2$ not
$\pi^0$,$\pi^+$,$\pi^-$ as three species of pions.
One can easily confirm that the calculation based on
$\pi^0$,$\pi^+$,$\pi^-$ leads to the same conclusion.

The factor attached to four point vertices depends on whether the
coupling concerns only one species of pions or not: $-\lambda$
for the $4\pi^0$ ($4\pi^1$ and $4\pi^2$)
vertices and $-\lambda/3$ for the $2\pi^0$-$2\pi^1$ ($2\pi^1$-$2\pi^2$
and $2\pi^2$-$2\pi^0$) vertices (see Fig.B.1.)

\bo{Fig.B.1}

First, let us take the $\pi^0+\pi^0\leftrightarrow 4\pi$ processes
with every possible combination of $4\pi$ such as
$\pi^++\pi^-+\pi^0+\pi^0$. All the possible Feynman diagrams are
shown in Fig.B.2. Since the three (two) diagrams in Fig.B.2(c) and
Fig.B.2(c${}^\prime$) (Fig.B.2(d)) have the same initial and final 
state, they must be added before taking the square and interfere 
with each other. Then we have to take possible combinations 
of allocating the external lines to the initial and final pions.
As is mentioned in Chap.2, we make the approximation that all the kinds of
phase space integrals give the same result. So, we just multiply the number
of combinations to the simplest diagrams shown in Fig.B.2 and then
put appropriate factors to avoid overcounting of the initial and final 
states. Now that phase space integrals are the same for all the
diagrams, we just need to calculate the factors coming from the
vertices ($-\lambda$ and $-\lambda/3$) and from the combination
of allocating the external lines shown above.

\bo{Fig.B.2}

These calculations are easily made and the diagrams in Fig.B.2
give ${3 \over 4}\lambda^4+{1\over 3}\lambda^4$ for
$\pi^1+\pi^1\leftrightarrow 4\pi$. The calculations for
$\pi^1+\pi^1\leftrightarrow 4\pi$ and $\pi^2+\pi^2\leftrightarrow
4\pi$ are made exactly in the same way, and one obtains
$3({3 \over 4}\lambda^4+{1\over 3}\lambda^4)$ as the sum of
$\pi^i+\pi^i\leftrightarrow 4\pi$ ($i=1,2,3$) reactions.

Next, we need to consider those reactions from (to) two different
species of pions to (from) all the possible four pions.
As an example, we take $\pi^0+\pi^1\leftrightarrow 4\pi.$
The possible processes are depicted in Fig.B.3. The calculations are
essentially the same as before, and we get ${56 \over 81}\lambda^4$
as the sum of these diagrams. The same results are obtained for the
other two processes $\pi^1+\pi^2\leftrightarrow 4\pi$ and
$\pi^2+\pi^0\leftrightarrow 4\pi$.

\bo{Fig.B.3}

Thus all the contribution of
isospin degrees of freedom  is $3({3\over 4}+{1\over 3}+{56\over
81})\lambda^4$, or the increase and decrease for each species of pion
is characterized by $({3\over 4}+{1\over 3}+{56\over 81})\lambda^4$.
In other word, the effect of isospin degrees of freedom can be taken
into account by replacing ${3\over 4}\lambda^4$, which is the factor
in the case of $\pi^0+\pi^0\leftrightarrow\pi^0+\pi^0+\pi^0+\pi^0$
given by Fig.B.2(a), by $({3\over 4}+{1\over 3}+{56\over 81})\lambda^4$.
The ratio $({3\over 4}+{1\over 3}+{56\over 81})\lambda^4 / 
{3\over 4}\lambda^4$ is the numerical factor $575\over 243$
in Eq.(\ref{ai}).

\newpage
\centerline{\LARGE Figure captions}
\begin{description}
\item{Fig.1:} The schematic picture  of heavy ion collision in $t$-$z$ space.

\item{Fig.2:}
%\begin{minipage}{0.8\textwidth}
The elementary process under consideration. (a):\scat\ scattering.
(b):\reac\ process. (c):the other possible \reac\ process in the second
order of perturbation in the $\sigma$ model. This diagram is
neglected. (d):All the possible detailed diagram of type (b).
(e):Approximated diagram for (d) used in the numerical calculation.
All the $4\pi$ vertices in (b),(c),(d),(e) are the effective
coupling obtained from the sum of diagrams in (a).
%\end{minipage}

\item{Fig.3.4.5.6:}
The time dependence of (a):temperature $T$, (b):chemical
potential $\mu$, and (c):$\log |\mu|$ for initial conditions
$(T_0,\mu_0)=(0.5,-1.0)$ (Fig.3),(1.5,-1) (Fig.4), (0.5,1) (Fig.5)
and (1.5,1.0) (Fig.6). The temperature $T$, chemical potential $\mu$,
time $t$ are all dimensionless ones defined in the text.

\item{Fig.7:} $\log_{10}{t_{\rm relax}}$ for various initial
conditions with $\lambda=120$.

\item{Fig.A.:} $x$ dependence of the temperature independent function $g(x)$.

\item{Fig.B.1:} Two kinds of $4\pi$ coupling in the $\sigma$ model.

\item{Fig.B.2:} All the possible Feynman diagrams for
$\pi^0+\pi^0\leftrightarrow 4\pi$

\item{Fig.B.2:} All the possible Feynman diagrams for
$\pi^0+\pi^1\leftrightarrow 4\pi$

\end{description}
\newpage 

%\begin{enumerate}

%\end{enumerate}

\bibb
\bib{lattice}{B.Petersson, Nucl.Phys.}{B}{(proc.Suppl.)30 (1993) 66
and references therin}
\bib{the}{T.Hatsuda, Nucl.Phys.}{A}{544 (1992) 27}
\bib{rev}{\it Quark Matter'84,}{}{Lecture Notes in Physics
221, Springer, New York, 1985}
\bib{mat}{T.Matsui et al., Phys.Rev.}{D}{34 (1986) 783}
\bib{berndt}{J.Rafelski and B.Mueller, Phys.Rev.Lett. 48}{}{(1984) 1066}
\bib{bertsch}{G.Bertsch et al., Phys.Rev.}{C}{37 (1988) 1896}
\bib{freeze}{L.V.Bravinu et al., Phys.Lett.}{B}{354 (1995) 196}
\bib{asa}{E.V.Shuryak, Sov.J.Nucl.Phys. 28}{}{(1978) 408}
\bib{landau}{L.D.Landau, Izv,Akad.Nauk SSSR}{17}{(1953) 51} 
\bib{deg}{DeGroot et al,}{}{\it Relativistic kinetic theory,\rm North Holland}
\bibe

\bye